\documentclass[doublecol,figures,amssymb,amsmath]{epl2}
\usepackage[latin1]{inputenc}
\usepackage[mathscr]{eucal}
\usepackage{bm}
\usepackage{amssymb}
\usepackage{amsmath}

\title{Measuring the distribution of current fluctuations through a Josephson junction with very short current pulses}
\shorttitle{Measuring current fluctuations through a JJ with very short pulses}

\author{Fabio Taddei\inst{1} \and Frank W. J. Hekking\inst{2}}
\shortauthor{F. Taddei and F. W. J. Hekking}

\institute{
  \inst{1} NEST CNR-INFM and Scuola Normale Superiore, I-56126 Pisa, Italy\\
  \inst{2} University Joseph Fourier and LPMMC-CNRS, B.P. 166, 25 Avenue des Martyrs, 38042 Grenoble-cedex 09, France
}

\abstract{We propose to probe the distribution of current
fluctuations by means of the escape probability histogram of a
Josephson junction (JJ), obtained using very short bias current
pulses in the adiabatic regime, where the low-frequency component
of the current fluctuations plays a crucial role. We analyze the
effect of the third cumulant on the histogram in the small
skewness limit, and address two concrete examples assuming
realistic parameters for the JJ. In the first one we study the
effects due to fluctuations produced by a tunnel junction, finding
that the signature of higher cumulants can be detected by taking
the derivative of the escape probability with respect to current.
In such a realistic situation, though, the determination of the
whole distribution of current fluctuations requires an
amplification of the cumulants. As a second example we consider
magnetic flux fluctuations acting on a SQUID produced by a random
telegraph source of noise.}

\begin{document}

\maketitle

\section{Introduction}
The electronic transport properties of a mesoscopic conductor are
completely characterized by its Full Counting Statistics (FCS),
introduced in Ref.~\cite{LLL} and defined as the probability
distribution for the transfer of charges over a certain time
interval. At zero temperature and in the absence of interaction,
for example, quantum transport is characterized by a binomial
distribution. This contrasts with the Poissonian statistics,
characteristic of classical uncorrelated particles, which is
recovered in the tunneling limit~\cite{reviewFCS}. The
experimental determination of FCS, unfortunately, is a difficult
task. Motivated by the expertise developed for current-noise
measurements, one possibility is to adopt the strategy of building
up the FCS through the measurement of the various cumulants of the
distribution, noise being the second one. The third cumulant was
indeed directly measured in Refs.~\cite{reulet,bomze05}. The
alternative possibility is the direct determination of the entire
FCS~\cite{nota1}.
Recently, various proposals have been put
forward in this direction making use of the high sensitivity to
fluctuations of current-biased Josephson junctions
(JJ)~\cite{tobiska,pekola04,heikkila04,ankerhold05,ankerhold07,pekola05,brosco06,timofeev,huard,suk}.
In this Letter we are interested in the second strategy and we
analyze the case in which the measurement is performed applying
very short current pulses, which allows to access a regime never
explored before.

According to the RCSJ model~\cite{tinkham}, the dynamics of the
phase difference across a current-biased JJ is equivalent to that
of a particle in a washboard potential, the phase playing the role
of the spatial coordinate. In the harmonic approximation, the
wells of such a tilted sine potential are characterized by the
plasma frequency $\omega_{\ab{p}}$. The wells are separated by
barriers if the current applied to the JJ is smaller than the
superconducting critical current $I_{\ab{c}}$. In such a case, the
phase-particle gets trapped in a well of the potential, causing
the voltage across the JJ to vanish (supercurrent state).
Escape from the well, which will cause the development of a finite
voltage across the JJ (transition to the resistive state), can
occur through two mechanisms. At low temperatures ($T\ll
\hbar\omega_{\ab{p}}/k_{\ab{B}}$), the only possibility is
macroscopic quantum tunneling (MQT) through the barrier which
separates two successive wells. The effect of thermal fluctuations
on the escape rate of the MQT was considered by Martinis and
Grabert~\cite{M+G}, who found an exponential enhancement of the
rate proportional to $T^2$ for ohmic damping. The second mechanism,
thermal activation (TA), is due to excitations of the particle
that will allow the latter to overcome the barrier top. In the
absence of perturbations, these are produced by thermal
fluctuations at large temperatures ($T>
\hbar\omega_{\ab{p}}/k_{\ab{B}}$).

In fact, the probability of escape from a potential well is
sensitive to perturbations affecting the system, such as
bias-current fluctuations. This is precisely the phenomenon
current fluctuations detection is based on. Indeed, by adding to
the bias current, assumed to be constant, the fluctuating
component of the current under investigation, it has been shown
that the monitoring of the voltage appearing across the JJ can be
used to characterize such fluctuations, either in the TA or MQT
regime. Regarding the former, in Ref.~\cite{tobiska} an array of
overdamped JJ which realizes a threshold detector was proposed,
whereas in Ref.~\cite{ankerhold07} a method was presented for the
determination of the third cumulant. On the other hand, MQT was
exploited for the determination of the FCS in Ref.~\cite{pekola04}
and of the fourth cumulant in Ref.~\cite{ankerhold05}. Successful
measurements in the TA and MQT regimes were very recently
reported, respectively, in Ref.~\cite{pekola05} for the second
cumulant, and in Refs.~\cite{timofeev,huard} for the third
cumulant. In this paper we are interested in hysteretic
(underdamped) JJ in the MQT regime. Inspired by the
experiments~\cite{claudon04,claudon06,claudon07}, we consider the
situation in which the measurement is performed by applying a
sequence of very short current pulses (of duration $\Delta t$ of
the order of a few nanoseconds)~\cite{nota}. This, together with a
proper filtering of the frequency components of the current
fluctuations, allows to realize a
separation of time scales, as we shall argue in the following (see
Fig.~\ref{system}(a)).

Current fluctuations whose frequency $\omega$ is larger than the
plasma frequency $\omega_{\ab{p}}$ give rise, even at zero
temperature, to thermal-like activation and represent the
non-adiabatic regime. For fluctuations of frequency smaller than
the plasma frequency, escape from the well can only occur through
MQT, which turns out to be exponentially sensitive to current
fluctuations. This is the adiabatic regime since the particle
remains in the ground state of the well. Within the adiabatic
regime, one can realize two different situations depending on
whether one selects the current fluctuations which occur on a time scale longer (low
frequency regime) or shorter (high frequency regime) than $\Delta
t$. For large $\Delta t$, of the order of $\mu$s, only the high
frequency (HF) regime is accessible, since long-time-scale
fluctuations do not contribute appreciably (see
Refs.~\cite{timofeev,pekola04}).
In this case the selection is obtained by 
placing a low-pass filter of bandwidth $\omega_{\ab{b}}\ll\omega_{\ab{p}}$ between the source of current fluctuations and the JJ.
In this Letter
we shall focus on the low-frequency
(LF) regime, where $\Delta t$ is of the order of a few
nanoseconds~\cite{claudon04,claudon06,claudon07}
and $\omega_{\ab{b}}\lesssim 2\pi/\Delta t\ll \omega_{\ab{p}}$ in order for the HF components not to affect the escape probability~\cite{shape}.
The current will fluctuate for different pulses, remaining constant within a single pulse.
We shall discuss
how the distribution of current fluctuations can be extracted from
the escape probability histogram and, in the small skewness limit,
how the former depends on the second and third cumulants. We shall
furthermore show that in some cases the influence of the current
distribution in the LF regime on the probability histogram is more
transparent than that relative to the HF regime. In addition we
discuss in detail two different examples.
We shall assume that the current fluctuations are produced by: i) a
tunnel junction; ii) a source of random telegraph noise~\cite{nn}.

\section{The system}
\begin{figure}
\onefigure[width=8.8cm]{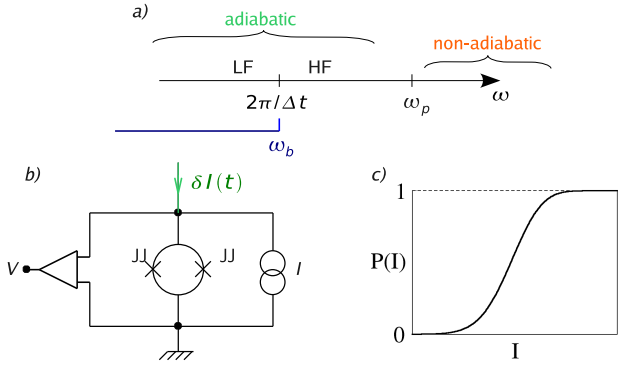} \caption{a) The frequency arrow:
fluctuations with frequency $\omega \ll \omega_{\ab{p}}$ produce
an adiabatic perturbation of the washboard potential, thus
affecting the escape probability with exponential sensitivity.
High frequency (HF) fluctuations ($\omega>2\pi/\Delta t$), occur over
the duration of a pulse giving rise to an averaged escape rate.
On the contrary, for low frequency (LF) fluctuations
 ($\omega<2\pi/\Delta t$), the current does not
fluctuate during a pulse, but changes from pulse to pulse giving
rise to an averaged escape probability.
In the LF regime we assume $\omega_{\ab{b}}\lesssim 2\pi/\Delta t\ll\omega_{\ab{p}}$, where $\omega_{\ab{b}}$ is the bandwidth.
Non-adiabatic perturbations are due to fluctuations with frequency
$\omega>\omega_{\ab{p}}$ which give rise to thermal-like
activation. b) The system consists of a current-biased dc-SQUID
connected to a voltage amplifier. $\delta I$ represents the
fluctuating component of the current flowing through a mesoscopic
element, whose current fluctuations distribution is under investigation.
c) Escape probability histogram obtained by varying
the bias current.}
\label{system}
\end{figure}

The system we consider (see Fig.~\ref{system}(b)) is a dc-SQUID,
composed of a superconducting loop with two identical JJ. We
consider a loop of small inductance, such that the SQUID is
equivalent to a single JJ with a flux-tunable critical current.
The SQUID is biased by a constant current source $I$ and coupled
to a voltage amplifier. The fluctuating component $\delta I(t)$ of
the current originating from a mesoscopic conductor is fed into
the SQUID through a proper filtering circuit (not shown in the
figure)~\cite{nnot}.
Such a circuit
represents the environment of the SQUID which must be taken into
account in the interpretation of the results of an experiment.
Indeed, the environment will introduce additional contributions to
the current fluctuations under investigation even exerting a
back-action on the system~\cite{beenakker03,kindermann04}. Since
the exact effect is determined by the details of the circuit, in
the following we shall disregard the effects of the environment
and focus only on the mere effects of current fluctuations.

Current fluctuations are characterized by the distribution $\rho
(\delta I)$ which can be written as a Fourier transform $\rho
(\delta I) = \frac{1}{2 \pi} \int_{- \infty}^{\infty} dk e^{- ik
\delta I} \phi_{\delta I} (k)$ of the characteristic
moment-generating function $\phi_{\delta I} (k) = \exp ( \sum_{n =
2}^{\infty} \frac{(ik)^n}{n!} c_n)$, where $c_n$ are the cumulants
of the distribution. In particular, $c_2$ is related to the width
($c_2 = \langle \delta I^2 \rangle$) and $c_3$ to the asymmetry of
the distribution ($c_3 = \langle \delta I^3 \rangle$).

The JJ, which is assumed to be underdamped, is operated as
follows. A sequence of current pulses, of duration $\Delta t$ and
amplitude $I$, is applied to the JJ, and the voltage across the JJ
is monitored. If, during the time interval $\Delta
t$, a finite voltage across the JJ develops, the event is
recorded. The ratio between the number of transitions and the
total number of pulses applied is defined as the escape
probability, which, as a function of the average amplitude $I$ of
the bias current, gives rise to a histogram $P(I)$ (see
Fig.~\ref{system}(c)). Such an escape probability histogram is
exponentially sensitive to perturbations affecting the system,
such as current fluctuations $\delta I(t)$ produced by the
mesoscopic element or fluctuations of the magnetic flux threading
the SQUID. We assume that temperature is small such that we can
neglect thermal fluctuations.

For very short pulses we access the LF regime, in which the current is constant within a single pulse, but fluctuates for different pulses.
Assuming ergodicity, the escape probability is
given by averaging the probability in the absence of fluctuations
$P_{\ab{0}}$ over the distribution of current fluctuations $\rho
(\delta I)$:
\begin{equation}
  P (I) = \int_{- \infty}^{+ \infty} d (\delta I) \rho (\delta
  I) P_{\ab{0}} (I + \delta I),
\label{pmeas}
\end{equation}
where $P_{\ab{0}} (I) = 1 - e^{- \Gamma \Delta t}$, $\Gamma =
\Gamma (I, \Phi)$ being the escape rate in the MQT regime given by
$\Gamma (I) \equiv A (I) \exp (- B (I))$, where $A (I) =
12\sqrt{\frac{6\pi \Delta U (I)}{\hbar\omega_{\ab{{p}}} (I)}}
\frac{\omega_{\ab{p}}(I)}{2\pi}$ and $B (I) = \frac{36}{5}
\frac{\Delta U (I)}{\hbar \omega_{\ab{{p}}} (I)}$ in the limit of
low dissipation~\cite{weiss}. The plasma frequency is given by
$\hbar \omega_{\ab{p}}=\sqrt{8E_{\ab{J}} E_{\ab{C}}}\left[ 2\left(
1-\frac{I}{I_{\ab{c}}} \right) \right]^{1/4}$, where
$E_{\ab{J}}=\hbar I_{\ab{c}}/2e$ is the Josephson energy,
$E_{\ab{C}}=e^2/2C_{\ab{J}}$ is the charging energy and
$I_{\ab{c}}$ is the critical current of the SQUID for zero
magnetic flux. $\Delta U$ denotes the barrier height and it is
defined as $\Delta U=\frac{2}{3}E_{\ab{J}}\left[ 2\left(
1-\frac{I}{I_{\ab{c}}} \right) \right]^{3/2}$.

The distribution $\rho (\delta I)$ can be derived directly by
deconvoluting the measured escape probability, once $P_0$ is
known. By Fourier transforming Eq.~(\ref{pmeas}) one obtains the
characteristic function
$\phi(k)=\tilde{P}(k)/\tilde{P}_0(k)$ where $\tilde{P}$
($\tilde{P}_0$) is the Fourier transform of $P$ ($P_0$). The
various cumulants of the distribution can then be calculated
through the derivatives
$
c_n=\left. (-i)^n\frac{\partial^n}{\partial k^n} \log \phi(k) \right|_{k=0} .
$
In the case where $P_{\ab{0}}$ has a very sharp transition
[$P_{\ab{0}} (I)\simeq \theta(I-I_0)$] the distribution can be
simply obtained by a derivation
\begin{equation}
  \rho (x)=\frac{dP(I_0-x)}{dx},
\label{sharp}
\end{equation}
so that the central moments of the distribution are obtained as
$\kappa_n=\int dx \; x^n \rho(x)$.
We wish to remark that in the LF regime the low frequency
component only will contribute to the moments and hence to $\rho$.
For example:
\begin{eqnarray}
\kappa_2=\int_{0}^{\frac{2\pi}{\Delta t}}\;\frac{d\omega}{2\pi}S_{II} (\omega)\\
\kappa_3=\int_{0}^{\frac{2\pi}{\Delta t}}\;\frac{d\omega_1}{2\pi}
\int_{0}^{\frac{2\pi}{\Delta t}}\;\frac{d\omega_2}{2\pi}S_{III}
(\omega_1,\omega_2) ,
\end{eqnarray}
where
\begin{eqnarray}
S_{II} (\omega)=\int_{-\infty}^{+\infty}dt\; e^{i\omega t} \langle \Delta I(t)\Delta I(0) \rangle\\
S_{III} (\omega_1,\omega_2)=\int_{-\infty}^{+\infty}dt_1\; e^{i\omega t_1}\int_{-\infty}^{+\infty}dt_2\; e^{i\omega t_2}\\
\nonumber\times \langle \Delta I(t_1) \Delta I(t_2)\Delta I(0) \rangle .
\end{eqnarray}
Higher frequency (adiabatic) components are filtered
out by the external circuit.

\section{Small skewness limit}
Neglecting the fourth and higher moments of current, and in the limit where
the skewness $\gamma$ of the current distribution is small ($\gamma \equiv c_3
/ c_2^{3 / 2} \ll 1$), one obtains
\begin{equation}
\rho (\delta I) \simeq \frac{1}{\sqrt{2 \pi c_2}} (1 -
  \frac{c_3}{2 c_2^2} \delta I + \frac{c_3}{6 c_2^3} \delta I^3) \exp (-
  \frac{\delta I^2}{2 c_2}) .
\label{small}
\end{equation}
We shall refer to the escape current $I_{\ab{esc}}$ as the current
relative to $P (I_{\ab{esc}})=1/2$, and to the width of the
transition as $\Delta I = I_{0.9} - I_{0.1}$, with $I_x$ defined
by $P (I_x) = x$.

\begin{figure}
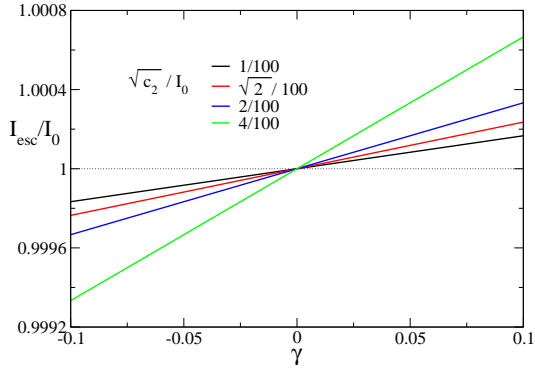

\onefigure[width=7cm]{2.eps} \caption{$I_{\ab{esc}}/I_0$ as a
function of $\gamma$ for several values of $c_2$. A linear
dependence is found with slope proportional to $\sqrt{c_2}/I_0$.}
\label{esck3}
\end{figure}

\begin{figure}
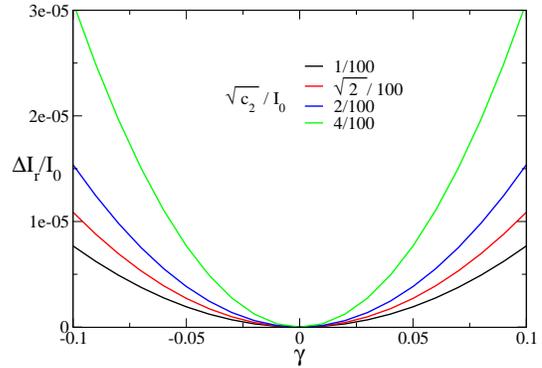

\onefigure[width=7cm]{3.eps} \caption{The normalized
relative width change $\Delta I_{\ab{r}} / I_0$ is plotted as a
function of $\gamma$ for several values of $c_2$.} \label{esck3di}
\end{figure}

For the sake of definiteness, we shall assume that $P_0$ has a
very sharp transition (with respect to the values of $c_2/I_0$
considered), so that we approximate the probability with a theta
function $P_0 (I)= \theta(I-I_0)$. Let us now discuss the behavior
of the probability histogram in the LF regime. For $c_3 = 0$
(Gaussian noise), the escape current does not depend on $c_2$,
while the width of the transition $\Delta I$ increases linearly
with $\sqrt{c_2}$. More precisely, $\Delta I=2\sqrt{2c_2}
\ab{Erf}^{-1}(4/5)$, where $\ab{Erf}^{-1}$ indicates the inverse
of the error function. Now, a finite skewness is expected to
produce an increase (decrease) of the escape current for $\gamma >
0$ ($\gamma < 0$). We calculate the escape probability and
determine $I_{\ab{esc}}$ and $\Delta I$ as a function of $\gamma$
for different values of $c_2$. As shown in Fig.~\ref{esck3},
$I_{\ab{esc}}$ increases linearly with $\gamma$, the slope being
proportional to $\sqrt{c_2}/I_0$ in such a way that
$(I_{\ab{esc}}-I_0)/\sqrt{c_2}=\ab{const}\cdot \gamma$. Since $\Delta
I$ at $\gamma=0$ depends on the value of $c_2$, in
Fig.~\ref{esck3di} we report the plot of the relative width change
$\Delta I_{\ab{r}}=\Delta I-\Delta I(\gamma=0)$ as a function of
$\gamma$. A quadratic dependence on $\gamma$ is found, with a
coefficient proportional to $\sqrt{c_2}/I_0$ so that $\Delta
I_{\ab{r}}/\sqrt{c_2}=\ab{const}\cdot \gamma^2$.
Note however that the dependence is very weak: 0.01\% escape
current change and 0.1\% transition width change in the range of
$\gamma$ considered. We wish to mention that the results in the
case of a finite-width transition, in the absence of fluctuations,
are qualitatively equal. For example, in the experimentally
relevant case in which $\Delta I_{\ab{int}}/I_0=5\times 10^{-3}$,
where $\Delta I_{\ab{int}}$ is the intrinsic width of $P_0$, we
find the same dependence of $I_{\ab{esc}}$ and $\Delta I$ on
$\gamma$. The only quantitative difference is that the values of
$\Delta I$ are shifted upwards, while the curve of $I_{\ab{esc}}$
is slightly shifted to lower values.

It is interesting to compare the above results with the ones
relative to the HF regime, in which the current fluctuates within
a single pulse. In such a case the escape probability is given by
averaging the escape rate over the noise distribution $\rho
(\delta I)$ of Eq.~(\ref{small}):
\begin{equation}
  P(I) = 1 - e^{- \langle \Gamma (I) \rangle \Delta t} = 1 -
  e^{- \int_{- \infty}^{+ \infty} d (\delta I) \rho (\delta I) \Gamma (I +
  \delta I) \Delta t} .
\end{equation}
It turns out that the width of the transition $\Delta I$ does not
depend on $\gamma$ so that the skewness affects the escape current
only.
However, since the influence of $c_2$ on the escape current
is strong, it is difficult to isolate the shift in the escape
current due to a change of $\gamma$.

In Ref.~\cite{peltonen} it was shown that, in the HF regime, the
effect of the third moment on the escape rate can be singled out
by inverting the polarity of the current $I$. Such inversion
results in a mere shift of the escape current which allows, in
principle, to evaluate the value of $\gamma$. We apply a similar
analysis to the LF regime finding, interestingly, that it is the
escape probability as a whole that is affected by the presence of
the third moment.
For the average escape probability
we obtain
\begin{eqnarray}
\label{asymmgP}\nonumber
P^{\pm} (I) = \frac{1}{\sqrt{2 \pi}} \int_{-
\infty}^{\infty} dxP_0 (I + \sqrt{c_2} x) \exp (- x^2
/ 2) \mp \\
\frac{\gamma}{2} \frac{1}{\sqrt{2 \pi}} \int_{- \infty}^{\infty}
dxP_0 (I + \sqrt{c_2} x) x (1 - x^2 / 3) \exp (- x^2 /2) ,
\end{eqnarray}
where the subscript $\pm$ refers to positive (negative) current
polarities. The values of $\gamma$ and $c_2$ can be determined if
$P_0$ is known. In particular, when $P_0$ can be approximated by a
theta function, the value of $\gamma$ can be determined from
\begin{equation}
P^+ (I) - P^- (I) = - \gamma \frac{e^{- \frac{(I
  - I_0)^2}{2 c_2}}}{2 c_2 \sqrt{2 \pi}} (I - I_0)^2,
\label{+-}
\end{equation}
while $c_2$ can be determined from the sum $P^+ +P^-$. It turns
out that the effect of the inversion of the current polarity is
smaller in the LF regime than in the HF regime. However, while for
LF $\gamma$ can be evaluated directly by comparing two histograms
(see Eq.~(\ref{+-})), for HF $\gamma$ can be determined only
indirectly from the escape rates and the histogram
derivative~\cite{peltonen}. Finally we note that spurious current
asymmetries may arise in the SQUID as a result of parasitic
magnetic fields. To avoid this phenomenon one should replace the
SQUID with a JJ as detector.

\section{Tunnel junction}
\begin{figure}
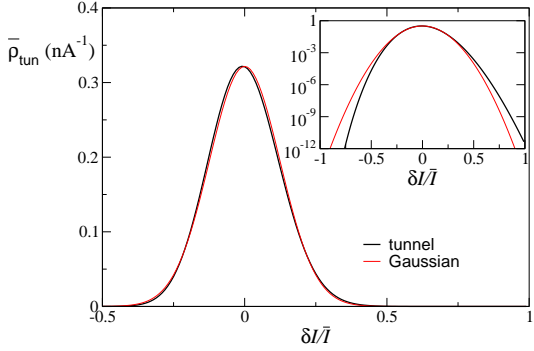

\onefigure[width=7cm]{4.eps}
\caption{Distribution of current fluctuations $\bar{\rho}_{\ab{tun}}(\delta I)$ relative to a tunnel
junction with $\Delta t=1$ ns and $\bar{I}=10$ nA (black curve) and of the corresponding Gaussian of
equal width (red curve). In the inset we plot the same curves in logarithmic scale.}
\label{rho}
\end{figure}

Let us consider a concrete example in which the mesoscopic element
is a tunnel junction. The current flowing through it is
distributed according to
 \begin{equation}
\rho_{\ab{tun}}(I)=\frac{T}{e}\frac{\bar{N}^{\frac{T}{e}I} e^{-\bar{N}}}{\Gamma \left( \frac{T}{e}I +1\right)} ,
\end{equation}
where $\bar{N}=T/e\bar{I}$ is the mean number of electron
transferred in a time $T$, $\bar{I}$ is the mean current, $e$ the
electronic charge, and $\Gamma$ is the Euler Gamma function.
Despite the fact that the FCS of a tunnel junction is Poissonian,
with all cumulants equal to $\bar{N}$, the cumulants of the
current distribution $\rho_{\ab{tun}}$ are given by
$c_n=\bar{I}(e/T)^{n-1}$, {\em i.e.} decreasing with increasing $n$. It is
interesting that for $\bar{N}$ as large as 100, $\rho_{\ab{tun}}$
is easily distinguishable from the Gaussian distribution of equal
$c_2$. Such small values of $\bar{N}$ can be achieved, in our
detection scheme, since the time $T$ can be identified with 
$2\pi/\omega_{\ab{b}}$, which is taken to be of the order of
the pulse duration $\Delta t$.
What limits the smallness of $\bar{N}$ is, firstly, the minimum achievable value $\Delta t$, and secondarily the maximum value of $\omega_{\ab{b}}$, which must be smaller than $\omega_{\ab{p}}$.
As an example, in Fig.~\ref{rho}, we
plot the function $\bar{\rho}_{\ab{tun}}(\delta
I)=\rho_{\ab{tun}}(\delta I+\bar{I})$ ({\em i. e.} the
distribution shifted by mean current $\bar{I}$) together with the
corresponding Gaussian (red curve), for $\Delta t=1$ ns and
$\bar{I}=10$ nA, so that $\bar{N}=65$ and
$\gamma=\bar{N}^{-1/2}\simeq 0.12$. For clarity, in the inset we
plot the same curves on logarithmic scale.

\begin{figure}
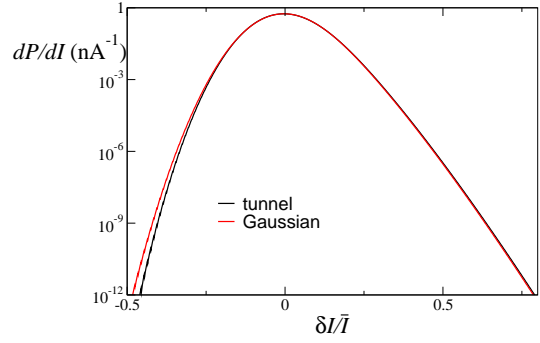

\onefigure[width=7cm]{5.eps} \caption{Logarithmic plot of the
derivative of $P_{\ab{tun}}$ relative to a tunnel junction (black
curve) for realistic parameters describing the JJ (see text). In
red is plotted the corresponding curve for a Gaussian current
distribution of equal width, which can be distinguished from the
black curve, in the range of values considered, at least for
negative fluctuations.} \label{der}
\end{figure}

The simplest way to measure $\rho_{\ab{tun}}$ in the LF regime is
through Eq.~(\ref{sharp}), valid in the limit of a step-like
transition in the escape histogram $P_0$. The latter condition
requires that the width of the transition in the absence of noise
$\Delta_{\ab{int}}$ is much smaller than the width of the
distribution $\sqrt{c_2}$. However, this is hardly possible with
realistic parameters, since the above two quantities are typically
of the same order. Indeed, in a typical experiment
$\Delta_{\ab{int}}/I_{\ab{c}}$ is of the order of $1\times
10^{-3}$, whereas with $\bar{I}=10$ nA, $I_{\ab{c}}=1$ $\mu$A and
$\Delta t=5$ ns one finds $\sqrt{c_2}/I_{\ab{c}}=6\times
10^{-4}$~\cite{numbers}. Surprisingly, the derivative of the
probability histogram $P_{\ab{tun}}$, obtained by
Eq.~(\ref{pmeas}) using $\rho_{\ab{tun}}$, turns out to be
distinguishable from the corresponding derivative of
$P_{\ab{Gauss}}$, relative to a Gaussian distribution of same
width. This can be appreciated in the left-hand-side of the plot
reported in Fig.~\ref{der}, where the above two quantities are
plotted as a function of the normalized current fluctuations. Even
in this far-from-ideal situation, the effect of higher cumulants
is weak but present~\cite{statis}. It is worthwhile mentioning
that, in order to make use of Eq.~(\ref{sharp}), one needs to
amplify the cumulants by a factor $\alpha$, for example through an
external circuit. For
the figures reported above, a value of $\alpha=100$ appears to be
sufficient for a determination of $\rho_{\ab{tun}}$.
We shall see in the next paragraph that the LF regime is much more
suited for peculiar distributions.

\section{Flux random telegraph noise}
\begin{figure}
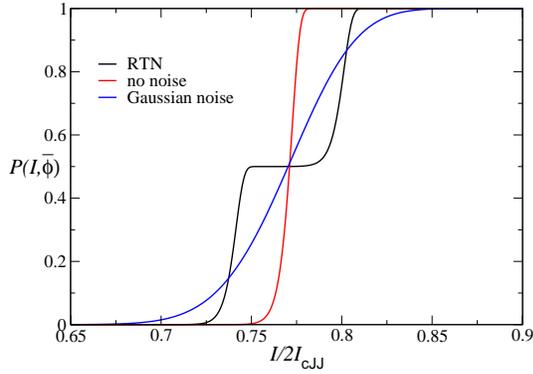

\onefigure[width=7cm]{6.eps}
\caption{Probability histograms with flux LF noise with $\bar{\phi}/\phi_0=1/5$, $\Delta\phi/\phi_0=1/60$, $I_{\ab{c}} = 1 \mu$A, $\Delta t= 5$ ns and $E_{\ab{C}} / E_{\ab{J}} = 10^{-4}$.
Black curve is relative to RTN, blue curve to Gaussian noise and red curve to no noise.
$I$ is expressed in units of $2 I_{\ab{cJJ}}$.}
\label{rtn1}
\end{figure}

Next we consider the situation where the flux threading the
dc-SQUID in Fig.~\ref{system} is affected by random telegraph
noise (RTN) and, for definiteness, where the current is noiseless.
The possibility of such a source of noise has been considered in
Ref.~\cite{tesiclaudon}. The time-dependent flux can be written as
$\phi (t) = \bar{\phi} + \Delta \phi \cos \left[ \pi \sum_k \theta
(t - t_k)\right]$, where $\bar{\phi}$ is the average flux, $\Delta
\phi$ is the displacement and $t_k$ are randomly distributed
times. The distribution of flux fluctuations $\delta\phi$ is given
by
\begin{equation}
  \rho_{\ab{flux}} (\delta\phi) = \frac{1}{2} \left[ \delta (\delta \phi -
  \Delta \phi) + \delta (\delta \phi + \Delta \phi) \right] ,
\end{equation}
with non-zero even central moments ($\langle (\delta \phi)^{2n}
\rangle = (\Delta \phi)^{2n}$) and characteristic function $\chi
(\lambda) = \frac{1}{2} \left( e^{i \lambda \Delta \phi} + e^{- i
  \lambda \Delta \phi} \right)$.
The time-dependent correlation function is given by $\langle
\delta \phi (t) \delta \phi (0) \rangle = \left( \Delta \phi
\right)^2 e^{- 2| t|p}$, where $p$ is the probability of
transition between the two states ($+ \Delta \phi / 2$, $- \Delta
\phi / 2$) per unit time. The noise power (double Fourier
transform) is therefore a Lorenzian $S (\omega) = \frac{(\Delta
\phi)^2}{2 \pi} \frac{2 p}{4 p^2 + \omega^2}$. In this case
one can choose the values of $p$ and $\Delta t$ such
that flux fluctuations has only a LF component. Indeed, if we
assume that $1/\Delta t \gg p$, the escape probability is given by
\begin{equation}
  P(I, \bar{\phi}) = \int d (\delta \phi) \rho_{\ab{flux}} (\delta
  \phi) P_0 (I, \bar{\phi} + \delta \phi),
\label{pifi}
\end{equation}
in the case where there is no current noise. In Eq.~(\ref{pifi}),
the flux dependence in $P_0 (I,\phi)$ is introduced by taking into
account the fact that $I_{\ab{c}}=2I_{\ab{cJJ}}\left| \cos\left(
\frac{\pi\phi}{\phi_0} \right) \right|$, where $I_{\ab{cJJ}}$ is
the critical current of a single JJ. In Fig.~\ref{rtn1},
$P(I,\bar{\phi})$ is plotted for $\bar{\phi}/\phi_0=1/5$,
$\Delta\phi/\phi_0=1/60$, $I_{\ab{c}} = 1 \mu$A, $\Delta t= 5$ ns
and $E_{\ab{C}} / E_{\ab{J}} = 10^{-4}$ ($\phi_0=h/2e$ being the
flux quantum). The red curve refers to the absence of noise, the
black curve to RTN and blue curve to Gaussian noise (of equal
width). The escape histogram relative to the RTN presents a double
step structure, well distinguishable from the histogram relative
to Gaussian noise. The double step histogram can be understood as
due to averaging two histograms with different escape current,
once a finite average flux $\bar{\phi}$ is assumed. However, the
effect of RTN noise in the HF regime leads to a decreased escape
current. The general feature of the HF regime is confirmed also in
this case: the effect of higher cumulants on the histogram reduces
to a shift in the escape probability, see also
Ref.~{\cite{ankerhold05}}.

\section{Conclusions}
To summarize, we have shown that the escape probability histogram
of an underdamped current-biased dc-SQUID, operated with very
short current pulses, can be used for retrieving information on
the distribution of current fluctuations which characterizes a
given conductor. The latter is connected to the dc-SQUID through
some filtering circuit, which we do not consider, for the sake of
definiteness, but would affect the measurement. More precisely,
this method is sensitive to the low frequency component of the
fluctuations, namely up to $1/\Delta t$ ($\Delta t$ being the
duration of the pulse, of the order of a few nanoseconds). We have
first analyzed the case in which the distribution presents a small
third moment (small skewness limit), while all other moments are
vanishing. In particular, we have addressed the functional
dependence of escape current and width of the transition with
respect to the skewness, and found that, even for short pulses,
the value of the skewness can be in principle evaluated by
inverting the polarity of the bias current. We have then analyzed
two concrete examples: in the first one we have assumed that the
current fluctuations originates from a tunnel junction and we have
evaluated the possibility of determining the whole current
distribution, by simulating the probability histogram under
realistic conditions.
We wish to mention that the same analysis can be performed
for any mesoscopic element.
In the second example, we have considered
flux fluctuations characterized by a random telegraph noise
distribution, finding that peculiar distributions such as the latter
produce a very clear signature, not present if the measurement is
performed with long current pulses.
Finally we wish to stress that we focused the analysis to
current pulses only, even though it remains valid for the case of
flux pulses.

\acknowledgments This work benefited from discussions with O.
Buisson, R. Fazio and J. P. Pekola. We also thank H. Grabert and
H. Pothier for a careful reading of the paper. F. T. acknowledges
the hospitality of CNRS, Grenoble. Partial financial support from
Institut Universitaire de France and the EU funded projects
NanoSciERA ``NanoFridge'', RTNNANO and EuroSQIP is acknowledged.


\begin{thebibliography}{0}

\bibitem{LLL}
\Name{Levitov L. S., Lee H. B. \and Lesovik G. B.}
\REVIEW{J. Math. Phys.}{37}{1996}{4845}.

\bibitem{reviewFCS}
  \Editor{Yu. V. Nazarov}
  \Book{Quantum Noise in Mesoscopic Physics}
  \Publ{Kluwer, Dordrecht}
  \Year{2003}.

\bibitem{reulet}
\Name{Reulet B., Senzier J. \and Prober D. E.}
\REVIEW{Phys. Rev. Lett.}{91}{2003}{196601}.

\bibitem{bomze05}
\Name{Bomze Yu., Gershon G., Shovkun D., Levitov L. S., \and Reznikov M.}
\REVIEW{Phys. Rev. Lett.}{95}{2005}{176601}.

\bibitem{nota1}
It is worthwhile mentioning that measurements of the whole FCS were reported in Refs.~\cite{gustavsson1,gustavsson2} for quantum dots (QDs) through a nearby quantum point contact (QPC). These experiments, however, are suitable only for QDs, since the QPC detector is sensitive to the charge present in the QD for a long enough time.

\bibitem{gustavsson1}
\Name{Gustavsson S., Leturcq R., Simovi\v c B., Schleser R., Ihn T., Studerus P., Ensslin K., Driscoll D. C. \and Gossard A. C.}
\REVIEW{Phys. Rev. Lett.}{96}{2006}{076605}.

\bibitem{gustavsson2}
\Name{Gustavsson S., Leturcq R., Simovi\v c B., Schleser R., Studerus P., Ihn T., Ensslin K., Driscoll D. C. \and Gossard A. C.}
\REVIEW{Phys. Rev. B}{74}{2006}{195305}.

\bibitem{tobiska}
\Name{Tobiska J. \and Nazarov Yu. V.}
\REVIEW{Phys. Rev. Lett.}{93}{2004}{106801}.

\bibitem{pekola04}
\Name{Pekola J. P.}
\REVIEW{Phys. Rev. Lett.}{93}{2004}{206601}.

\bibitem{heikkila04}
\Name{Heikkila T. T., Virtanen P., Johansson G. \and Wilhelm F. K.}
\REVIEW{Phys. Rev. Lett.}{93}{2004}{247005}.

\bibitem{ankerhold05}
\Name{Ankerhold J. \and Grabert H.}
\REVIEW{Phys. Rev. Lett.}{95}{2005}{186601}.

\bibitem{ankerhold07}
\Name{Ankerhold J.}
\REVIEW{Phys. Rev. Lett}{98}{2007}{036601}.

\bibitem{pekola05}
\Name{Pekola J. P., Nieminen T. E., Meschke M., Kivioja J. M., Niskanen A. O. \and Vartiainen J. J.}
\REVIEW{Phys. Rev. Lett.}{95}{2005}{197004}.

\bibitem{brosco06}
\Name{Brosco V., Fazio R. , Hekking F. W. J. \and Pekola J. P.}
\REVIEW{Phys. Rev. B}{74}{2006}{024524}.

\bibitem{timofeev}
\Name{Timofeev A. V., Meschke M., Peltonen J. T., Heikkil\"a T. T. \and Pekola J. P.}
\REVIEW{Phys. Rev. Lett.}{98}{2007}{207001}.

\bibitem{huard}
\Name{Huard B., Pothier H., Birge N.O., Est\`eve D., Waintal X.
\and Ankerhold, J.} arXiv:0711.0646.

\bibitem{suk}
\Name{Sukhorukov V. \and Jordan A. N.}
\REVIEW{Phys. Rev. Lett.}{98}{2007}{136803}.

\bibitem{tinkham}
  \Name{Tinkham M.}
  \Book{Introduction to Superconductivity}
  \Publ{Dover Publications, New York}
  \Year{1996}.

\bibitem{M+G}
\Name{J. M. Martinis, and H. Grabert} \REVIEW{Phys. Rev.
B}{38}{1988}{2371}.

\bibitem{claudon04}
\Name{Claudon J., Balestro F., Hekking F. W. J. \and Buisson O.}
\REVIEW{Phys. Rev. Lett.}{93}{2004}{187003}.

\bibitem{claudon06}
\Name{Claudon J., Fay A., L\'evy L. P. \and Buisson O.}
\REVIEW{Phys. Rev. B}{73}{2006}{180502(R)}.

\bibitem{claudon07}
\Name{Claudon J., Fay A., Hoskinson E. \and Buisson O.}
\REVIEW{Phys. Rev. B}{76}{2007}{024508}.

\bibitem{nota}
To be precise, in Refs.~\cite{claudon04,claudon06,claudon07} nanosecond-long {\em flux} pulses were used.

\bibitem{shape}
We assume that the shape of the pulses is such that adiabaticity
is preserved: rise and fall times should be long enough so that
their inverses are smaller than $\omega_{\ab{p}}$. This point was
addressed in the experiment reported in Ref.~\cite{claudon07}, where nanosecond pulses were used.

\bibitem{nn}
For instance induced by a fluctuating magnetic flux threading the weak link.

\bibitem{nnot}
A possible circuit is a $LC$ low-pass filter, setting $\omega_{\ab{b}}=1/\sqrt{LC}$. For $\omega\lesssim\omega_{\ab{b}}$, the overall Q factor of the JJ is virtually unaffected. See also Ref.~\cite{peltonen}.

\bibitem{peltonen}
\Name{Peltonen J. T., Timofeev A. V., Meschke M. \and Pekola J. P.}
\REVIEW{J. Low Temp. Phys.}{146}{2007}{135}.

\bibitem{beenakker03}
\Name{Beenakker C. W. J., Kindermann M. \and Nazarov Yu. V.}
\REVIEW{Phys. Rev. Lett.}{90}{2003}{176802}.

\bibitem{kindermann04}
\Name{Kindermann M., Nazarov Yu. V. \and Beenakker C. W. J.}
\REVIEW{Phys. Rev. B}{69}{2004}{035336}.


\bibitem{weiss}
  \Name{Weiss U.}
  \Book{Quantum Dissipative Systems}
  \Publ{World Scientific, Singapore}
  \Year{1999}.

\bibitem{numbers}
We assume $E_{\ab{C}}/E_{\ab{J}}\simeq 1\times 10^{-7}$, which
yields a value of $\Delta I_{\ab{int}}/I_{\ab{c}}$ compatible with
the experimental one.

\bibitem{statis}
Note that the difference between the two curves can be appreciated
mainly for rare events. For example we find that
$P_{\ab{tun}}(-0.5)=8\times10^{-8}$, meaning that, to collect
enough statistics, one needs a number of pulses of the order of
$10^{9}$.

\bibitem{tesiclaudon}
  \Name{Claudon J.}
  \Book{Oscillations coh\'erentes dans un circuit quantique supraconducteur: le SQUID dc}
  \Publ{Universit\'e Joseph Fourier, Grenoble}
  \Year{2005}.



\end{thebibliography}
\end{document}